\documentclass[twocolumn,pra,superscriptaddress]{revtex4-1}
\usepackage[utf8]{inputenc}
\usepackage{float}
\usepackage{placeins}
\usepackage{qcircuit}
\usepackage{braket}
\usepackage{amsmath}
\usepackage[normalem]{ulem}
\usepackage{todonotes}
\usepackage{xcolor}
\usepackage{multirow}
\usepackage{braket}
\usepackage{bm}
\usepackage{bbold}
\usepackage{todonotes}
\usepackage{wrapfig}
\usepackage{appendix}
\usepackage{graphicx,lipsum}% http://ctan.org/pkg/{graphicx,lipsum}
\newcommand{\PRLsep}{\noindent\makebox[\linewidth]{\resizebox{0.65\linewidth}{1.5pt}{$\bullet$}}\bigskip}

\newcommand{\ketbra}[2]{\mbox{$|#1\rangle\langle #2|$}}

\def\ketbra#1#2{{\vert#1\rangle\!\langle#2\vert}}

\def\braket#1#2{{\langle#1\vert#2\rangle}}

\usepackage{mathtools} % Bonus
\graphicspath{{figures/}}

\begin{document}

\title{Robustness of Variational Quantum Algorithms against stochastic parameter perturbation}

\author{Daniil Rabinovich}\affiliation{Skolkovo Institute of Science and
Technology, Moscow, Russian Federation}
\affiliation{Moscow Institute of Physics and Technology,  Moscow, Russian Federation}
\author{Ernesto Campos}\affiliation{Skolkovo Institute of Science and
Technology, Moscow, Russian Federation}
\author{Soumik Adhikary}\affiliation{Skolkovo Institute of Science and
Technology, Moscow, Russian Federation}
\author{Ekaterina Pankovets}\affiliation{Skolkovo Institute of Science and Technology, Moscow, Russian Federation}
\affiliation{Moscow Institute of Physics and Technology,  Moscow, Russian Federation}
\author{Dmitry Vinichenko}\affiliation{Skolkovo Institute of Science and
Technology, Moscow, Russian Federation}
\affiliation{Moscow Engineering Physics Institute, Moscow, Russian Federation}
\author{Jacob Biamonte}\affiliation{Beijing Institute of Mathematical Sciences and Applications, Beijing, China}
% \homepage{http://quantum.skoltech.ru}

\begin{abstract}
Variational quantum algorithms are tailored to perform within the constraints of current quantum devices, yet they are limited by performance-degrading errors. In this study, we consider a noise model that reflects realistic gate errors inherent to variational quantum algorithms. We investigate the decoherence of a variationally prepared quantum state due to this noise model, which causes a deviation from the energy estimation in the variational approach. By performing a perturbative analysis of optimized circuits, we determine the noise threshold at which the criteria set by the stability lemma is met. We assess our findings against the variational quantum eigensolver and quantum approximate optimization algorithm for various problems with up to $14$ qubits. Moreover, we show that certain gate errors have a significantly smaller impact on the coherence of the state, allowing us to reduce the execution time without compromising performance.

% Variational algorithms are designed to work within the limitations of contemporary devices and suffer from performance limiting errors. Here we identify an experimentally relevant model for gate errors, natural to variational quantum algorithms. We study how a quantum state prepared variationally decoheres under this noise model, which manifests as a perturbation to the energy approximation in the variational paradigm. A perturbative analysis of an optimized circuit allows us to determine the noise threshold for which the acceptance criteria imposed by the stability lemma remains satisfied. We benchmark the results against the variational quantum approximate optimization algorithm for $3$-SAT instances and unstructured search with up to $10$ qubits and $30$ layers. Finally, we observe that errors in certain gates have a significantly smaller impact on the quality of the prepared state. Motivated by this, we show that it is possible to reduce the execution time of the algorithm with minimal to no impact on the performance. 
\end{abstract}

 \maketitle

 \section{Introduction}

    Noisy Intermediate Scale Quantum (NISQ) computing \cite{preskill2018} is constrained by limited coherence times and operation precision \cite{weidenfeller2022scaling, ratcliffe2018scaling, hegde2022toward, mills2022two}, which restrict the number of qubits and circuit depths that can be implemented with reasonable fidelity. This limits the range of possible experimental demonstrations. 
    % However, a growing body of theoretical research is improving our understanding of the use of random circuit sampling as a basis for a scalable experimental violation of the extended Church-Turing thesis \cite{AGL+22} and on the complexity analysis of NISQ \cite{CCL+22} algorithms.
    The variational model of quantum computation is tailored to operate within these practical limitations \cite{kandala2017hardware, Rabinovich22, pagano2020}, and has been shown to be computationally universal under idealized conditions \cite{biamonte2021universal}. 
    % Nevertheless, it is still susceptible to certain types of noise.
    % Reminiscent of machine learning, a variational algorithm makes use of a short parameterized quantum circuit, known as ansatz, in which parameters are iteratively tuned to minimize a cost function in a quantum-to-classical feedback loop \cite{cerezo2021variational}. The cost function is typically given in the form of the expectation of a so called problem Hamiltonian; where the ground state of the problem Hamiltonian encodes the solution of a given problem instance. Thus, by the way of cost function (energy) minimization, a variational algorithm attempts to approximate the ground state of a given Hamiltonian. This strategy, however, does not provide us with a guarantee in regards to the quality of the approximate solution, where the latter is typically quantified as the overlap between the state prepared by the ansatz and the true ground state. Nevertheless, the overlap can be bounded. It has been shown using the stability lemma that the bounds can be directly related to the energy, thus allowing us to determine the energy threshold (upper bound) required to guarantee a fixed minimum overlap. We call this the {\it acceptance} threshold; a state with energy below this threshold is said to be accepted by the algorithm \cite{biamonte2021universal}. 
    Similar to machine learning, a variational algorithm employs a parameterized quantum circuit, called an ansatz, that is iteratively adjusted to minimize a cost function in a quantum-to-classical feedback loop \cite{cerezo2021variational}. The cost function usually takes the form of the expectation of a problem Hamiltonian, where the ground state of the problem Hamiltonian represents the solution to a given problem instance. By minimizing the cost function (energy), a variational algorithm aims to approximate the ground state of the Hamiltonian. However, this approach does not guarantee the quality of the approximate solution, which is typically measured by the overlap between the state prepared by the ansatz and the true ground state. Nonetheless, the overlap can be bounded. Using the stability lemma \cite{biamonte2021universal}, it has been demonstrated that the bounds can be directly linked to the energy, allowing us to determine the energy threshold (upper bound) required to ensure a minimum (fixed) overlap. We refer to this as the acceptance threshold, and a state with an energy below this threshold is considered accepted by the algorithm.

    Variational algorithms are designed to mitigate some of the systematic limitations of NISQ devices \cite{Harrigan2021, pagano2020, Guerreschi2019, Butko2020}. However, these algorithms are still susceptible to stochastic noise. While there is some evidence that variational algorithms can benefit from a certain level of stochastic noise \cite{Campos2021}, in general, noise negatively impacts their performance by inducing decoherence and degrading solution quality.

In this paper, we investigate how errors in the form of parameter deviations impact the performance of variational algorithms when operated at their noiseless optimal parameters. We analytically demonstrate that the energy shift varies quadratically with the spread of parameter deviation, equivalent to an energy shift linear with respect to the gate error probabilities for various noise models \cite{Dalton2022}. We validate our findings introducing the noise to the Variational Quantum Eigensolver (VQE) employed for Ising Hamiltonian, and the Quantum Approximate Optimization Algorithm (QAOA) for $3$-SAT \cite{Akshay2020}, MAX-CUT \cite{farhi2014quantum} and unstructured search \cite{akshay2021parameter, grover1996fast}. Furthermore, we observe the performance to be more resilient to alterations in certain parameters. Based on these findings, we propose methods to potentially enhance performance and reduce the execution time of variational quantum algorithms.

\nopagebreak
 \section{Preliminaries}
%  \subsection{VQE}
% Finding the ground state of a quantum system has long been identified as an application where quantum computers can have a tangible advantage \cite{Lloyd96}. Variational quantum eigensolvers (VQE) are purpose built to accomplish this task. The algorithm has three steps. Given a problem Hamiltonian $H$ we first prepare a variational state $\ket{\psi(\boldsymbol{\theta})} = U(\boldsymbol{\theta}) \ket{0}^{\otimes n} = \Big(\prod_{j=1}^p U_j (\theta_j) \Big)\ket{0}^{\otimes n}$. Here $U(\boldsymbol{\theta})$ is a variational ansatz and $\boldsymbol{\theta} \in [0, 2\pi)^{\times p}$ are tunable parameters. Next, the cost fucntion $\braket{\psi(\boldsymbol{\theta})|H}{\psi(\boldsymbol{\theta})}$ is constructed based on local measurements. In the final step the cost function is minimized using a classical co-processor which outputs: 
% \begin{eqnarray}
%     &\boldsymbol{\theta}^* \in \arg \min_{ \boldsymbol{\theta}} \braket{\psi(\boldsymbol{\theta})|H}{\psi(\boldsymbol{\theta})} \\
%    \nonumber \\
%     &E^* = \min_{ \boldsymbol{\theta}} \braket{\psi(\boldsymbol{\theta})|H}{\psi(\boldsymbol{\theta})} \\
%     \nonumber \\
%     &\ket{\psi(\boldsymbol{\theta}^*)} = U(\boldsymbol{\theta}^*) \ket{0}^{\otimes n}
% \end{eqnarray}

% Here, $\ket{\psi(\boldsymbol{\theta}^*)}$ is the approximate ground state of $H$. 

\subsection{Variational Quantum Eigensolver}
VQE is a variational algorithm designed to approximate the ground state and energy of a given $n$-qubit problem Hamiltonian $H$. The approximation to the ground state is prepared using a tunable  quantum circuit (a.k.a. ansatz) $U(\bm\theta)$ as
$\ket{\psi(\bm\theta)} = U(\bm\theta) \ket{0}^{\otimes n} = \Big(\prod_{k} U_k (\theta_k) \Big)\ket{0}^{\otimes n}$ with tunable parameters $\theta_k\in [0, 2\pi)$. 
Upon the preparation of quantum state $\ket{\psi(\bm\theta)}$, local measurements  are used to calculate the energy $E(\bm\theta)=\bra{\psi(\bm\theta)} H \ket{\psi(\bm\theta)}$. Finally, a classical co-processor iteratively tunes variational parameters to minimize the energy function, to identify 
\begin{eqnarray}
\label{eq:E^*}
    &\bm\theta^* \in \arg \min_{ \bm\theta} E(\bm\theta), \\
   \nonumber \\
   \label{eq:theta^*}
    &E^* = \min_{ \bm\theta} E(\bm\theta),
    % \nonumber \\
    % &\ket{\psi(\theta^*)} = U(\theta^*) \ket{0}^{\otimes n},
\end{eqnarray}
which gives an approximation to the ground energy of $H$.

A certain type of VQE, called Quantum Approximate Optimization Algorithm (QAOA) \cite{farhi2014quantum} was specifically designed to approximately solve combinatorial optimization problems \cite{niu2019optimizing,farhi2014quantum,lloyd2018quantum,morales2020universality,Zhou2020,wang2020x,Brady2021,Farhi2016,Akshay2020,Farhi2019a,Wauters2020,Claes2021,Zhou}. In this algorithm the combinatorial problem is encoded into a diagonal problem Hamiltonian $H$, whose ground state encodes the solution to the original problem. % The algorithm employs ansatze circuits expressive enough to (in theory) emulate any quantum cirucuit \cite{lloyd2018quantum, morales2020universality}. 
 % Consider a pseudo-Boolean function $\mathcal{C}: \{0,1\}^{\times n} \rightarrow \mathbb{R}$, the objective of the algorithm is to approximate a bit string that minimizes $\mathcal{C}$. To accomplish this, $\mathcal{C}$ is first encoded as a problem Hamiltonian $H$, diagonal in the computational basis. The ground state of $H$ encodes the solution to the problem; in other words QAOA searches for a solution $\ket{g}$ such that $\braket{g \vert H}{g} = \min H$. 
 The algorithm makes use of a problem-dependent ansatz of depth $p$, which prepares a variational state
$\ket{\psi_p (\bm\gamma, \bm\beta)}$ as
\begin{equation}
    \ket{\psi_p(\bm\gamma,\bm\beta)} =  \prod\limits_{k=1}^p e^{-i \beta_k H_{x}} e^{-i \gamma_k H}\ket{+}^{\otimes{n}},
    \label{ansatz}
\end{equation} 
with real parameters $\gamma_k\in[0,2\pi)$, $\beta_k\in[0,\pi)$. Here $H_x = \sum_{j=1}^{n} X_{j}$ is the standard one-body mixer Hamiltonian with Pauli  matrix $X_j$ applied to the $j$-th qubit. Similar to standard VQE, the algorithm minimizes the energy cost function $E(\bm\gamma, \bm\beta) = \bra{\psi_p(\bm\gamma,\bm\beta)}H \ket{\psi_p(\bm\gamma,\bm\beta)}$ as in \eqref{eq:E^*} and \eqref{eq:theta^*} with $\bm\theta = (\bm\gamma, \bm\beta)$.
% \begin{eqnarray}
%     \label{eq:approx_eng}
%     &E^*=\min_{\bm\gamma,\bm\beta} \bra{\psi_p(\bm\gamma,\bm\beta)}H \ket{\psi_p(\bm\gamma,\bm\beta)},\\
%     \nonumber \\
%     \label{eq:approx_sol}
%     &{\bm \gamma}^*, {\bm \beta}^* \in \arg \min_{\bm\gamma,\bm\beta} \bra{\psi_p(\bm\gamma,\bm\beta)}H \ket{\psi_p(\bm\gamma,\bm\beta)}
% \end{eqnarray}
The state $\ket{\psi_p({\bm\gamma}^*,{\bm\beta}^*)}$ approximates the ground state of $H$ and hence encodes the solution to the original problem. Nevertheless, in any VQE the quality of the approximation, quantified as the overlap between the true solution and the approximate solution, is not known a priori from  \eqref{eq:E^*}. Still, one can establish bounds on this quantity using the so called stability lemma. 

\subsection{Stability lemma}

The stability lemma states that if $\ket{g}$ is the true ground state of $H$ with energy $E_g$ and $\Delta$ is the spectral gap (the difference between the ground state energy and the energy of the first excited state) the following relation holds \cite{biamonte2021universal, akshay22}:
\begin{equation}
    1 - \frac{E^*- E_g}{\Delta} \leq \vert \braket{\psi_p({\bm\gamma}^*,{\bm\beta}^*)}{g} \vert^2 \leq 1 - \frac{E^*-E_g}{E_m-E_g}
    \label{eq:stability}
\end{equation}
where $E_m$ is the maximum eigenvalue of $H$. Thus to guarantee a non-trivial overlap one must ensure that $E^* \leq E_g + \Delta$. We call the latter the acceptance condition.

\section{Quantum circuits in the presence of realistic gate errors}
\label{sec:noise_model}
Implementation of unitary operations used in variational ansatze depends significantly on the considered hardware. Typically the implementation makes
use of electromagnetic pulses, such as in superconducting quantum computers \cite{clarke08,gambetta17}, neutral atom based quantum computers \cite{Gerasimov2022,Morgado2021}, and trapped ion based quantum computers \cite{pagano2020,Zhang2017}. Such pulses can change the population of the energy levels that constitute a qubit or introduce phases to the quantum amplitudes, thus controlling the state of the qubits. 
Fluctuations in the pulse shape, phase and amplitude will affect the operation and reduce its fidelity \cite{brown2021materials,schindler2013quantum}. 
As in typical unitary operations the angles of rotations depend on time averaged intensity $I(t)$ of the electromagnetic pulse ($\theta \propto \int I(t) dt$),  variations in pulse parameters can result in stochastic deviations of the angles from the desired values, which becomes increasingly important for fast gate realizations \cite{gale2020optimized}. 
Moreover, these angle deviations can be present even in the case of perfectly stabilized pulses: in ion based quantum computers thermal motion reduces certainty in ions positions, leading to variations in the ion-laser couplings \cite{brown2021materials,benhelm2008towards} and causing deviations in the resulting angles \cite{gale2020optimized}.
This type of error can contribute gate infidelity  of about $10^{-3}-10^{-2}$ \cite{benhelm2008towards, monroe2008quantum}; while other sources of noise would also contribute to gate imperfections, the discussed model becomes the dominant source of error for certain setups, such as ground state ion qubits with radial modes \cite{Egan2021}.

In such settings if a quantum circuit is composed of the parameterised gates $\{U_k(\theta_k)\}_{k=1}^q$; ${\theta_k \in [0, 2\pi)}$ and one tries to prepare a state $\ket{\psi(\bm\theta)}=\prod_{k=1}^q U_{k}(\theta_k)\ket{\psi_0}$, a different state 
\begin{equation}
    \ket{\psi(\bm \theta+\bm{\delta\theta})}=\prod\limits_{k=1}^{q}U(\theta_k+\delta\theta_k)\ket{\psi_0},
\end{equation}
is prepared instead due to the presence of errors.
Notice here that the perturbation $\bm{\delta\theta}$ to the parameters is stochastic and is sampled with a certain probability density $p(\bm{\delta\theta})$. This implies that the prepared state can be described by an ensemble $\{\ket{\psi(\bm\theta+\bm{\delta\theta})},p(\bm{\delta\theta})\}$, which can be treated as a density matrix
 \begin{equation}
    \rho(\bm\theta)=\int\limits_{\bm{\delta\theta}\in[-\pi,\pi]^{\times q}} p(\bm{\delta\theta})\ketbra{\psi(\bm\theta+\bm{\delta\theta})}{\psi(\bm\theta+\bm{\delta\theta})}d(\bm{\delta\theta}).
    \label{eq:dm}
\end{equation}

Eq.~\eqref{eq:dm} represents a noise model native to the variational paradigm of quantum computing. Thus, for the rest of the paper we systematically study the effect of this noise model on the performance of Variational Quantum Algorithms, specifically VQE and QAOA. In particular we study the energy perturbation around $E^*$ in different scenarios subsequently recovering the strength of noise under which the  acceptance condition continues to be satisfied.

% We show that the behaviour of the perturbation can be analytically predicted for small amounts of noise. We subsequently confirm this numerically and recover the strength of noise under which the condition of acceptance continues to be satisfied. 
\newpage
\section{Energy of perturbed quantum circuits}
\label{sec:results}
In this section we present our analytical and numerical findings on the robustness of VQE against the noise model discussed in Section \ref{sec:noise_model}. Specifically, we study how energy $E = \bra{\psi(\bm\theta)}H\ket{\psi(\bm\theta)}$ of a prepared variational state $\ket{\psi(\bm\theta)}$ changes when $\bm\theta$ is perturbed. In our analysis we do not require $\bm\theta$ to be an optimum of the energy function, as in \eqref{eq:E^*}. 
Nevertheless, following the settings of VQE, we apply our analysis to optimized circuits $U(\bm\theta^*)$, which prepare the best possible approximations to the ground state for a given circuit depth. Finally, we substantiate our findings with numerical simulations of the perturbed optimized circuits.

\subsection{Perturbative analysis in presence of gate errors}
\label{sec:analysis_pert}

Consider a problem Hamiltonian $H$ and a variational ansatz $\ket{\psi(\bm\theta)}=U_1(\theta_1)\dots U_q(\theta_q)\ket{\psi_0}$ used to minimize~$H$. Here the gates $U_k(\theta_k)$ have the form:
\begin{equation}
\label{eq:op}
    U_k(\theta_k)=e^{iA_k\theta_k}, A_k^2=\mathbb{1}.
\end{equation}

A typical example of such an ansatz is the checkerboard ansatz \cite{Uvarov2021}, with M$\o{}$lmer-S$\o{}$rensen (MS) gates as the entangling two qubit gates. Nevertheless, any quantum circuit can admit a decomposition in terms of operations of the form \eqref{eq:op}; this adds generality to this assumption.

In the presence of gate errors the prepared quantum state decoheres as $\ket{\psi(\bm\theta)} \rightarrow \rho(\bm\theta)$ as per \eqref{eq:dm}.
To obtain the analytic form of $\rho(\bm\theta)$ we first note that 

\begin{widetext}
\begin{align}
U_k(\theta_k+\delta\theta_k)=U_k(\theta_k)U_k(\delta\theta_k)\nonumber=\cos\delta\theta_k U_k(\theta_k)+\sin\delta\theta_k U_k\left(\theta_k+\frac{\pi}{2}\right).
\end{align}
This follows directly from \eqref{eq:op}. Therefore we get:

\begin{align}
\label{eq:dm_projector}
&\ketbra{\psi(\bm\theta+\bm{\delta\theta})}{\psi(\bm\theta+\bm{\delta\theta})}=\sum\limits_{k_1,\dots,k_q,m_1,\dots,m_q=0}^1(\cos^{2}\delta\theta_1\tan^{k_1+m_1}\delta\theta_1)\dots(\cos^{2}\delta\theta_q\tan^{k_q+m_q}\delta\theta_q)\ketbra{\psi_{k_1\dots k_q}}{\psi_{m_1\dots m_q}}, 
\end{align}
\end{widetext}
where 
\begin{equation}
    \ket{\psi_{k_1\dots k_q}(\bm \theta)}=U_1\Big(\theta_1+k_1\frac{\pi}{2}\Big)\dots U_q\Big(\theta_q+k_q\frac{\pi}{2}\Big)\ket{\psi_0}.
\end{equation}

Here we make three realistic assumptions---(a) perturbations to all the angles are independent, (b) the distribution is symmetric around zero, $p(\bm{\delta\theta})=p(-\bm{\delta\theta})$  and (c) the distribution $p(\delta\theta_k)$ vanishes quickly outside the range $(-\sigma_k,\sigma_k)$; that is, the error is localized on the scale $\sigma_k\ll1$. Note that if assumption (b) does not hold, as long as mean value $\langle\delta\theta_k\rangle$ is independent from the angle $\theta_k$, one can always shift the parameters as $\theta_k\to\theta_k-\langle\delta\theta_k\rangle$ to avoid non-zero mean.
Otherwise, terms linear in $\langle\delta\theta_k\rangle$ could contribute to the energy perturbation \cite{Kattemolle2022}. Notice that we do not require $\delta\theta_k$ to be small compared to the value $\theta_k$.

% Should it not be the case, the shift $\langle\delta\theta\rangle$ can be included into optimal parameters, $\theta^*\to\theta^*+\langle\delta\theta\rangle$, and then mean value of a new perturbation vanishes. We also assume that the induced error is localized on the scale $\sigma_k\ll1$, i.e. $p(\delta\theta_k)$ quickly vanishes outside the range $(-\sigma_k,\sigma_k)$. For example, for normally distributed perturbation $\delta\theta_k$, $\sigma_k$ will be the variance of the distribution. 

Substituting \eqref{eq:dm_projector} in \eqref{eq:dm} we arrive at the expression:

\begin{equation}
\label{eq:rho}
    \rho (\bm \theta)=\ketbra{\psi(\bm \theta)}{\psi(\bm \theta)}+\delta\rho,
\end{equation}
where

\begin{equation}
\label{eq:delta_rho}
    \delta\rho\approx -\sum\limits_{k=1}^qa_k\ketbra{\psi(\bm \theta)}{\psi(\bm \theta)} + \sum_{k=1}^q a_k\ketbra{\psi_{k}}{\psi_{k}}+o(\sigma_k^2).
\end{equation}

% up to terms that are quadratic in $\sigma_k$. 
Here $\ket{\psi_k}\equiv\ket{\psi_k(\bm \theta)} = \ket{\psi_{00...1...00}(\bm \theta)}$ with $1$ placed in the $k$-th position. We point out that the terms of the form $\ketbra{\psi_{k}}{\psi_{m}}$ for $k\ne m$ do not enter \eqref{eq:delta_rho} as the associated factor $\langle \delta \theta_k \delta\theta_m\rangle =\langle \delta \theta_k\rangle\langle\delta\theta_m\rangle = 0$. Finally, 

\begin{equation}
\label{eq:a_k}
    a_k\equiv\langle\sin^2\delta\theta_k\rangle\approx\langle\delta\theta_k^2\rangle=\int(\delta\theta_k)^2 p(\delta\theta_k)d(\delta\theta_k)\sim \sigma_k^2.
\end{equation}
The numerical prefactor in the last transition depends on the exact form of the distribution and we neglect it for generality. 
% We introduce $a_k\equiv\langle\sin^2\delta\theta_k\rangle=\int\sin^2\delta\theta_k p(\delta\theta_k)d(\delta\theta_k)\sim \sigma_k^2$, $\langle\cos^2\delta\theta_k\rangle=1-\langle\sin^2\delta\theta_k\rangle=1-a_k$. We also notice that $\langle\sin\delta\theta_k\rangle\sim \sigma_k^3\ll a_k$, which follows from the Taylor expansion and assumption (b). 
% Thus, from \eqref{eq:dm} and \eqref{dm_projector} up to terms quadratic in $\sigma_k$ we establish 
% \begin{align}
% \rho=\ketbra{\psi_0}{\psi_0}+\delta\rho\\
% \delta\rho\approx -\sum\limits_{k=1}^qa_k\ketbra{\psi_0}{\psi_0} + a_1\ketbra{\psi_{10\dots0}}{\psi_{10\dots0}}+\nonumber\\
% a_2\ketbra{\psi_{01\dots0}}{\psi_{01\dots}}+\dots+a_q\ketbra{\psi_{00\dots1}}{\psi_{00\dots1}}.
% \label{dm_correction}
% \end{align}

% The final step in \eqref{eq:a_k} becomes possible due to the assumption (c).
Notice that \eqref{eq:delta_rho} can be viewed as the action of certain noisy channel, where each of the gates is altered with probability $a_k\sim \sigma_k^2$. In this sense, we call $a_k$ gate error probabilities, though this treatment is specific to the interpretation of the noisy channel.

For the simplest case of identical distributions ($\sigma_k^2=\sigma^2=a$), the noise induced energy perturbation around the noiseless value $E = \bra{\psi(\bm\theta)}H\ket{\psi(\bm\theta)}$ becomes
 \begin{align}
     \delta E \equiv  {\rm Tr}(\rho(\bm\theta) H)-E&= {\rm Tr}(\delta\rho H)\nonumber\\
     &\approx a\sum\limits_{k=1}^{q} (\bra{\psi_k}H\ket{\psi_k}-E),
     \label{eq:delta E}
 \end{align}
which demonstrates that energy perturbation depends linearly on the gate error probabilities $a$ (quadratic in $\sigma$). A similar dependence on the gate error probabilities was observed earlier \cite{Dalton2022}, albeit with a different noise model.

In the most general setting one cannot estimate energies $\bra{\psi_k}H\ket{\psi_k}$, as they strongly depend on the problem Hamiltonian, considered ansatz and its parameters.
Nevertheless, by imposing a trivial bound $\bra{\psi_k}H\ket{\psi_k}\leq E_m$ we can establish
\begin{equation}
\delta E\leq qa(E_m-E)\sim q\sigma^2(E_m-E).
\label{eq:E_sigma}
\end{equation}
% For the practical purposes of VQE, where the ansatz state attempts to approximate the ground state, following the minimization \eqref{eq:E^*} we can estimate amount of noise, tolerated by stability lemma.  Namely, from 
Now we turn to practical applications of VQE, where the ansatz attempts to approximate the ground state of the problem Hamiltonian. Following the optimization \eqref{eq:E^*}, noiseless energy $E$ becomes $E^*$---the energy of the optimized circuit, which in the presence of noise increases by $\delta E$ according to \eqref{eq:E_sigma}. According to the stability lemma \eqref{eq:stability}, we request that noisy energy $E^*+\delta E< E_g+\Delta$ and conclude that for 
\begin{equation}
    \sigma<\sqrt{\dfrac{\Delta-(E^*-E_g)}{q(E_m-E^*)}}
    \label{eq:sigma_bound}
\end{equation}
the acceptance condition is still satisfied.
In other words, this level of noise can be tolerated by the algorithm, as it still can produce states with certain overlap with the ground state. Obviously, this requires the original noiseless circuit to prepare a good approximation of the ground state, i.e. $E^*<E_g+\Delta$.
In practice, however, the bound \eqref{eq:sigma_bound} can be significantly relaxed with tighter bounds on $\bra{\psi_k}H\ket{\psi_k}$, possible only for specific problems. 

When performing the approximation in \eqref{eq:delta E} we disregarded higher powers of $\sigma_k$. In the most general case this is justified when $qa\sim q\sigma^2\ll 1$, which can further be improved for specific problems, where terms like $\bra{\psi_k}H\ket{\psi_k}$ could be estimated.

\subsection{Numerical simulations of stochastic perturbations}
While our perturbative analysis holds for all VQEs, we substantiate our findings by numerically minimizing instances of Ising Hamiltonian with checkerbord ansatz and implementing QAOA for a range of problems. 
% In particular we solve instances of 3-SAT and unstructured search problems to study the behaviour of energy perturbation around $E^*$ caused by the presence of gate errors.
Importantly, we numerically confirm validity of the quadratic approximation \eqref{eq:delta E} all the way up to $q\sigma^2\gtrsim 1$ for specific problems, significantly expanding applicability range of treatment compared toanalytical predictions. Indeed, for realistic infidelities of NISQ devices ($a\sim0.001-0.01$) this could allow for execution of thousands of gates. 

In our numerical simulations we start by minimizing (finding the best approximation for the considered depth) specific problem Hamiltonians with chosen ansatz to identify optimal set of angles $\theta^*$ ($(\bm\gamma^*, \bm\beta^*)$ for QAOA). 
%Notice that optimization does not imply nearly perfect ground state preparation, but rather the best possible one for the considered circuit depth. 
Then the optimal circuit is subjected to the noise model described by \eqref{eq:dm}; for that we randomly sample perturbations $\delta$ to each of the gate angles from a uniform distribution on the interval $(-\sigma,\sigma)$ and average the obtained energies.

\subsubsection{VQE for Ising Hamiltonian}
\begin{figure}[ht!]
    \centering
    \includegraphics[width=0.49\textwidth]{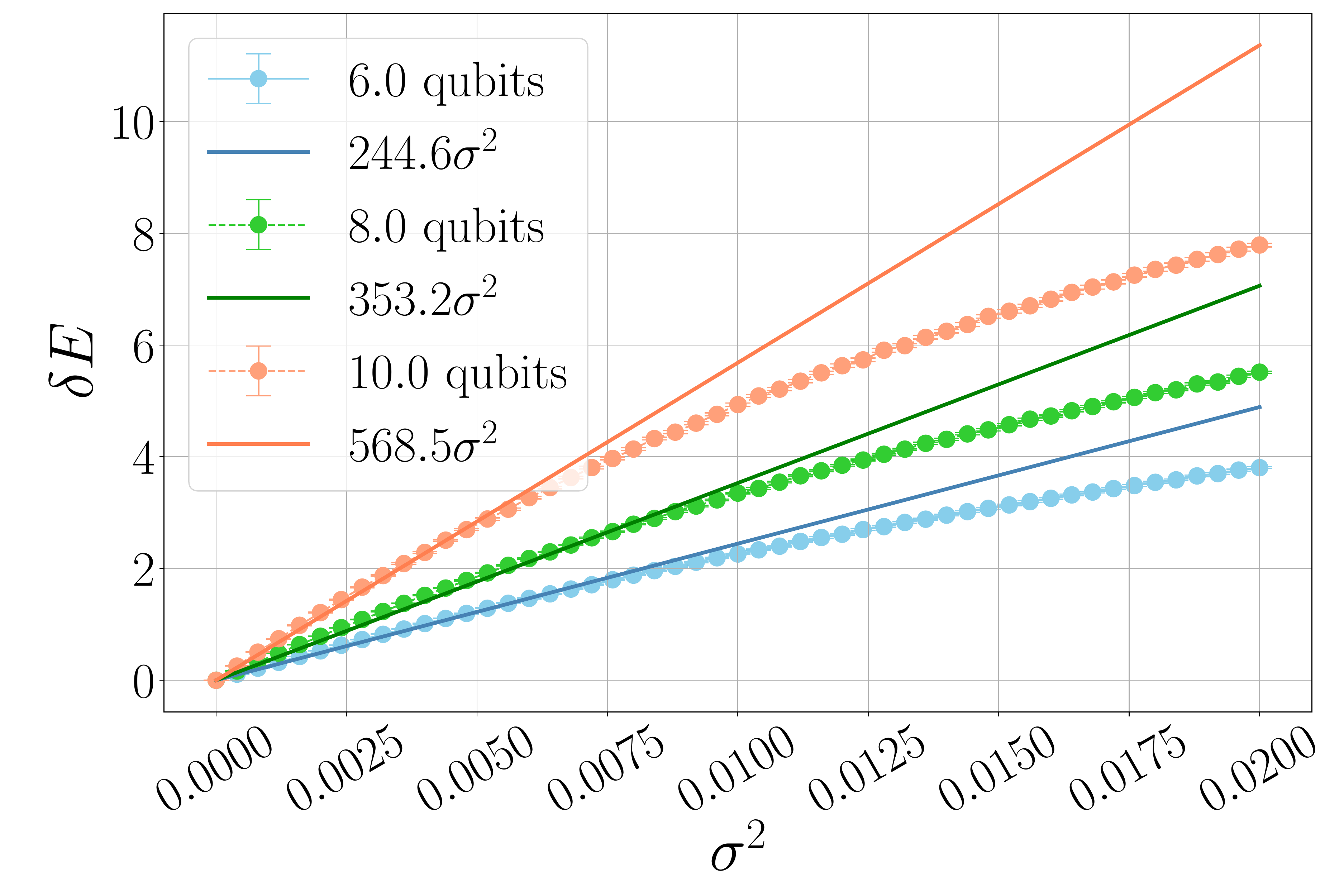}
    \caption{Average energy increase of perturbed optimal VQE circuit of $n=6, 8, 10$ qubits for circuit depth $p=8,10,12$,  obtained by the perturbation of $\bm\theta^*$ by $\delta$ uniformly sampled from the range $(-\sigma,\sigma)$. Error bars depict standard errors.
    Polynomial fits of data points confirm that $\delta E \propto \sigma^2$.}
    \label{fig:VQE}
\end{figure}

We begin by considering minimization of Ising Hamiltonian
\begin{equation}
\label{eq:Ising}
    H = \sum_{j=1}^n Z_j Z_{j+1} + h \sum_{j=1}^n X_j,
\end{equation}
with the checkerboard ansatz, composed of Molmer-Sorensen ($R_{XX}$) \cite{Sorensen1999, Sorensen2000} entangling gates and local rotations. 
Here we set $Z_1 = Z_{n+1}$. We consider $n=6, 8, 10$ with circuit depth $p=8, 10, 12$ respectively, which, after optimization, guaranteed more than $50\%$ overlap with the ground state. For every $n$ we considered $100$ instances of Hamiltonian \eqref{eq:Ising} with transverse field in the range $h\in[0.8,1.2]$. 
Subjecting every circuit to the discussed noise model, we average the energy perturbation over considered instances and obtain Fig. \ref{fig:VQE}.

Indeed, it confirms our analytical prediction of quadratic scaling of $\delta E$ with respect to $\sigma$ up to $\sigma^2\sim10^{-2}$. Moreover, it can be seen that proportionality coefficient indeed is roughly proportional to number of gates in checkerbord ansatz $q = 5np$, as expected from $\eqref{eq:E_sigma}$. Importantly, the linear dependence prevails for $\sigma$ up to $q\sigma^2\gtrsim 1$ ($q = 240, 400, 600$ for $n=6, 8, 10$ respectively), significantly expanding applicability of our result.

\subsubsection{QAOA}
In this section we focus on the effect of perturbing QAOA circuits, optimized to solve instances of various problems. We begin by considering 3-SAT problem, defined in appendix \ref{appen:problems}. 
Importantly, in our numerical implementation the problem Hamiltonian propagator $e^{-i\gamma H}$ is realized as a sequence of gates of the type \eqref{eq:op} (notice that $H$ is naturally written in the Pauli basis, which gives the desired decomposition) to adhere to practical realizations. Angles in all these gates receive uncorrelated perturbations; the same holds for the single qubit gates implementing $e^{-i \beta H_x}$.

\begin{figure}[ht!]
    \centering
    \includegraphics[width=0.49\textwidth]{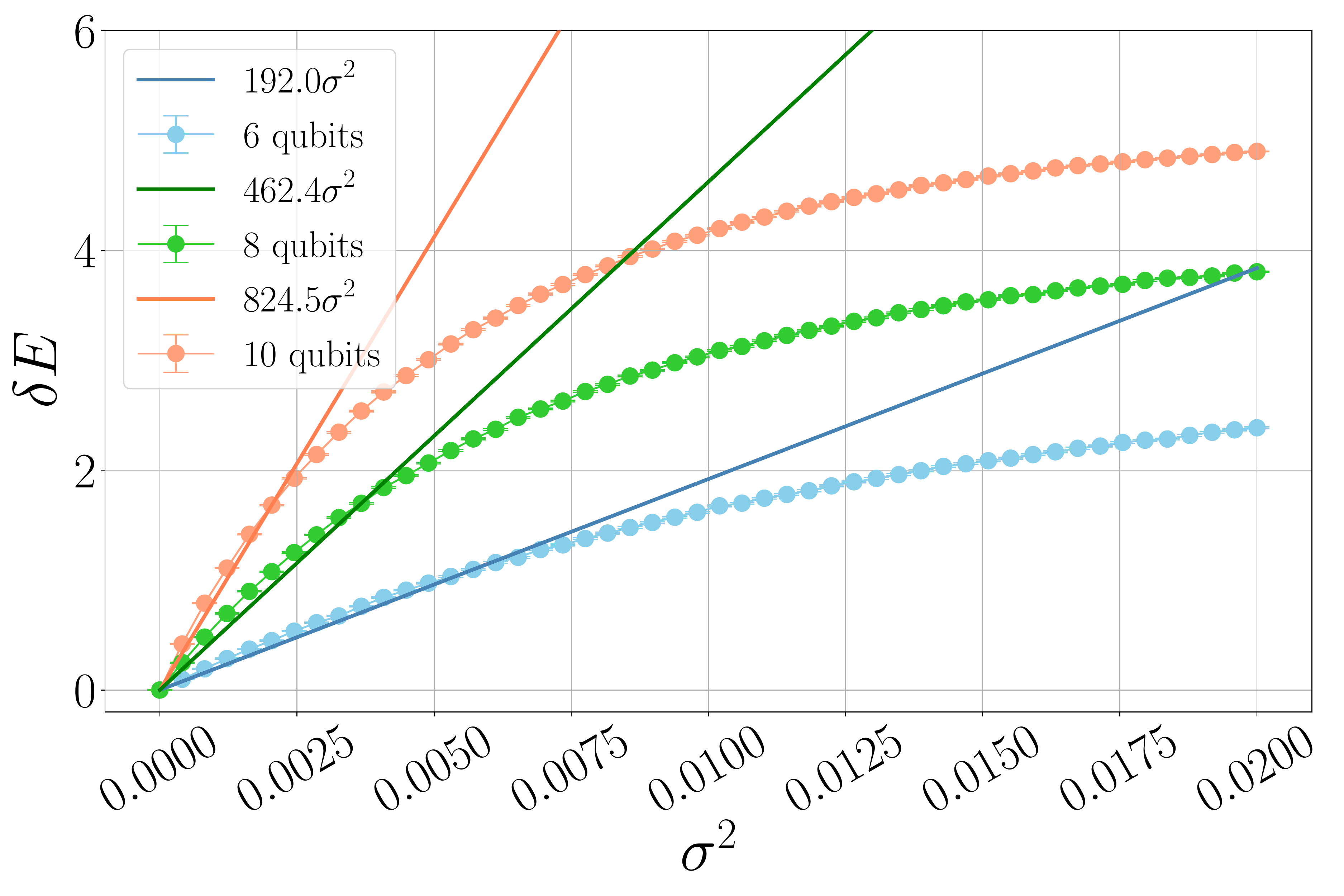}
    \caption{Average energy shift of 100 uniformly generated 3-SAT instances of 6, 8 and 10 qubits obtained from perturbed optimal circuits of depth $15$, $25$ and $30$,  respectively. All instances were designed to have clause to variable ratio of 4.2 and a unique satisfying assignment. 
    % The shifts are obtained  by the perturbation of  $\bm \gamma^*,~ \bm\beta^*$ by $\delta$ uniformly sampled from the range $(-\sigma,\sigma)$. 
    Error bars depict standard errorss. Polynomial fits of data indicate $\delta E\propto \sigma^2$ up to $q\sigma^2\gtrsim5$.}
    \label{fig:averaged_pert}
\end{figure}

We ran QAOA for 100 uniformly generated 3-SAT instances of 6, 8, and 10 variables with 26,
34 and 42 clauses (corresponding to computational phase transition density of $4.2$ for 3-SAT) and circuits of $15$, $25$ and $30$ layers, respectively. All the instances were generated to have a unique satisfying assignment. The optimal circuits are then perturbed and the calculated energies are averaged over considered instances. 
The results are presented in Fig. \ref{fig:averaged_pert}. 
Again, it is seen that for small values of noise the energy scales as $\delta E \propto \sigma^2$, 
as per \eqref{eq:E_sigma}, which is equivalent to linear dependence on the gate error probabilities $a_k$. 
The linear behaviour is now limited to a smaller range of $\sigma$, compared to the case depicted in Fig. \ref{fig:VQE}. This is caused by a larger gate count in the QAOA circuits (primarily from the decomposition of $e^{-i\gamma H}$): number of gates, averaged over the instances, reach $q = 555, 1525, 2610$ for $n=6, 8, 10$ respectively. Nevertheless, the linear approximation remains valid up to $q\sigma^2\gtrsim 5$, again improving over analytical predictions.
% It is seen, that the value $\sigma\sim0.075$ could never violate the acceptance criteria, as corresponding  energy error never exceeds the gap $\Delta\ge1$. 
% For smaller number of qubits and gates the threshold value of $\sigma$ increases. 

% \todo[inline]{Figure 3: applicability domain shrinks. Needed depth falls out of NISQ ranges, leading to large energies, causing the deviation from a line. Thus, depth has to be bounded; thus, the new plot, where considered depth is within limitations of NISQ devises.}

\begin{figure}[ht!]
    \centering
    \includegraphics[width=0.49\textwidth]{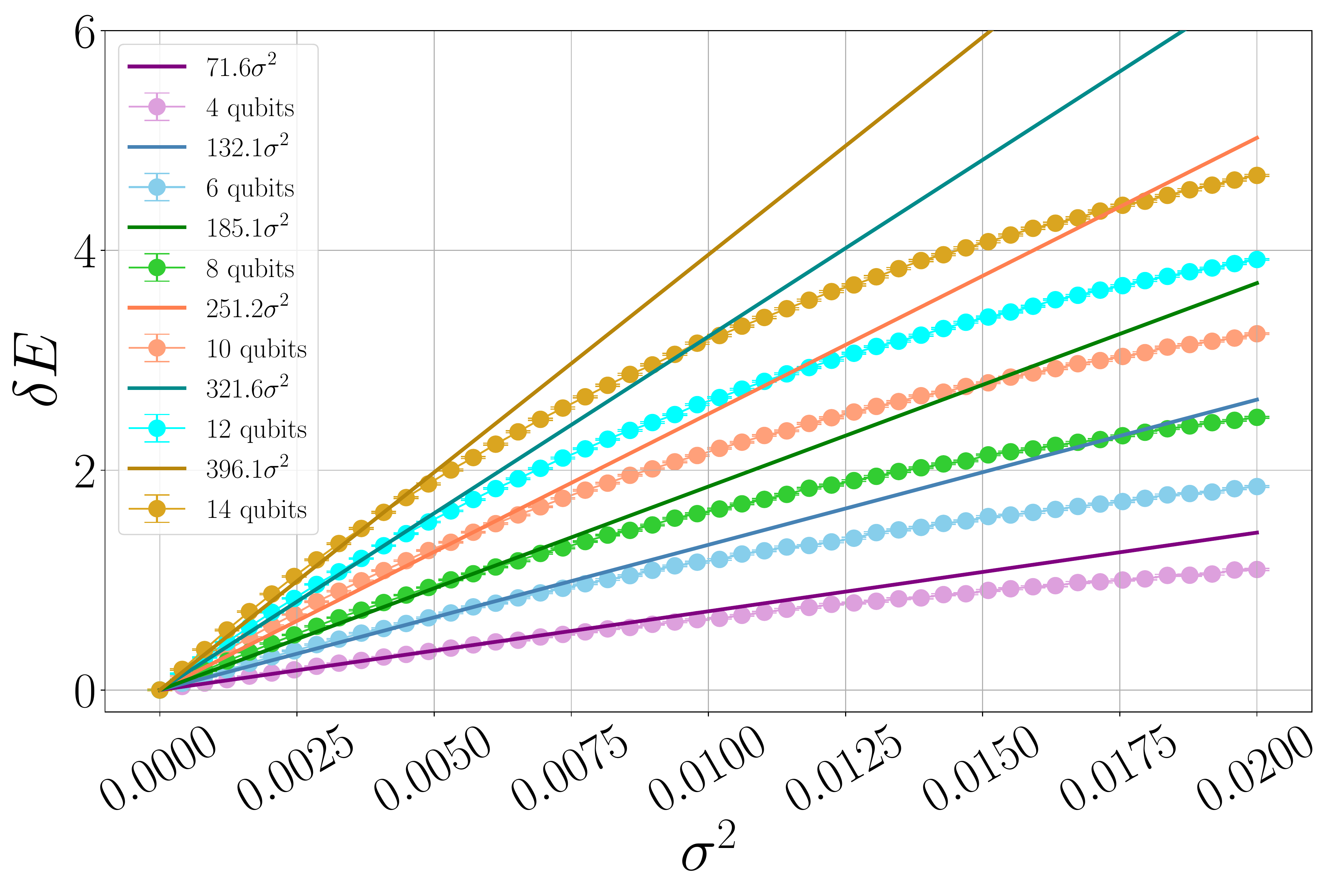}
    \caption{Average energy shift of 100 uniformly generated 3-SAT instances of 6, 8 and 10 qubits obtained from perturbed optimal circuits of depth $10$. All instances were designed to have clause to variable ratio of 4.2 and a unique satisfying assignment. 
    Error bars depict standard errors. Polynomial fits of data indicate $\delta E\propto \sigma^2$ up to $q\sigma^2\gtrsim5$.}
    \label{fig:p=10}
\end{figure}

These large gate counts $q$, induced by the circuit depth sufficient to solve 3-SAT, might already be beyond the capabilities of current NISQ devices. However, if one were interested in a decent approximate solution to the considered problem, the required circuit depth could be reduced, allowing to meet NISQ realities. Indeed, we obtained decent approximate solutions to instances of 3-SAT with a fixed circuit depth $p=10$, which constituted less than $1000$ gates for up to $n=10$ qubits, reaching at most $1500$ gates for $n=14$ qubits. By perturbing these circuit, we see that energy increase, depicted in Fig. \ref{fig:p=10}, is better approximated by a linear dependence with respect to gate errors, compared to Fig. \ref{fig:averaged_pert}.

Additionally, in Appendix \ref{appen:other_results} we present results for QAOA solving instances of MAX-CUT and unstructured search (defined in Appendix \ref{appen:problems}). Importantly, for the latter we considered the case of highly correlated pertubations (without decomposition of $e^{-i\gamma H}$ into gates of the form \eqref{eq:op}) to demonstrate generalizability of our results. Both cases qualitatively agree with results obtained for VQE and MAX-SAT. Overall, based on the presented evidence we conclude that linear scaling of energy perturbation with respect to $\sigma^2$ prevails all the way up to $q\sigma^2\gtrsim 1$. Thus, for the infidelities of current devices ($\sigma^2\lesssim0.001-0.01$) our prediction remains valid for circuits composed of thousands of gates.  Importantly, actual values of energies deviate towards lower values, thus making our prediction a valid upper bound on energy perturbations even for larger number of gates and noise levels.
% \begin{figure}[ht!]
%     \centering
%     \includegraphics[width=0.49\textwidth]{figures/H1_all_sigma2.pdf}
%     \caption{Average energy for the problem of unstructured search obtained by the perturbation of $\bm \gamma^*,~ \bm\beta^*$ by $\delta$ uniformly sampled from the range $(-\sigma,\sigma)$. Error bars depict standard errors. Polynomial fits of data points of 6, 8 and 10 qubits in the ranges $\sigma \in [0,0.1], ~[0,0.07], ~[0,0.05]$, respectively confirm that $\delta E \propto \sigma^2$.}
%     \label{fig:state_prep_sigma}
% \end{figure}

% For unstructured search,  we average the energy over $\delta$ sampled for each gate from the uniform distribution $(-\sigma,\sigma)$. We again recover that $\delta E\propto\sigma^2$, as depicted in Fig.~\ref{fig:state_prep_sigma}.
% It is seen that the same threshold $\sigma\sim 0.075$ now increases energy by no more then $0.6$, which guaranties $40\%$ overlap with the target state. 

\subsection{Perturbation to individual parameters}
\begin{figure}[!tbh]
 \begin{minipage}[b]{\linewidth}
   \raggedright{\includegraphics[clip=true,width=3.05in]{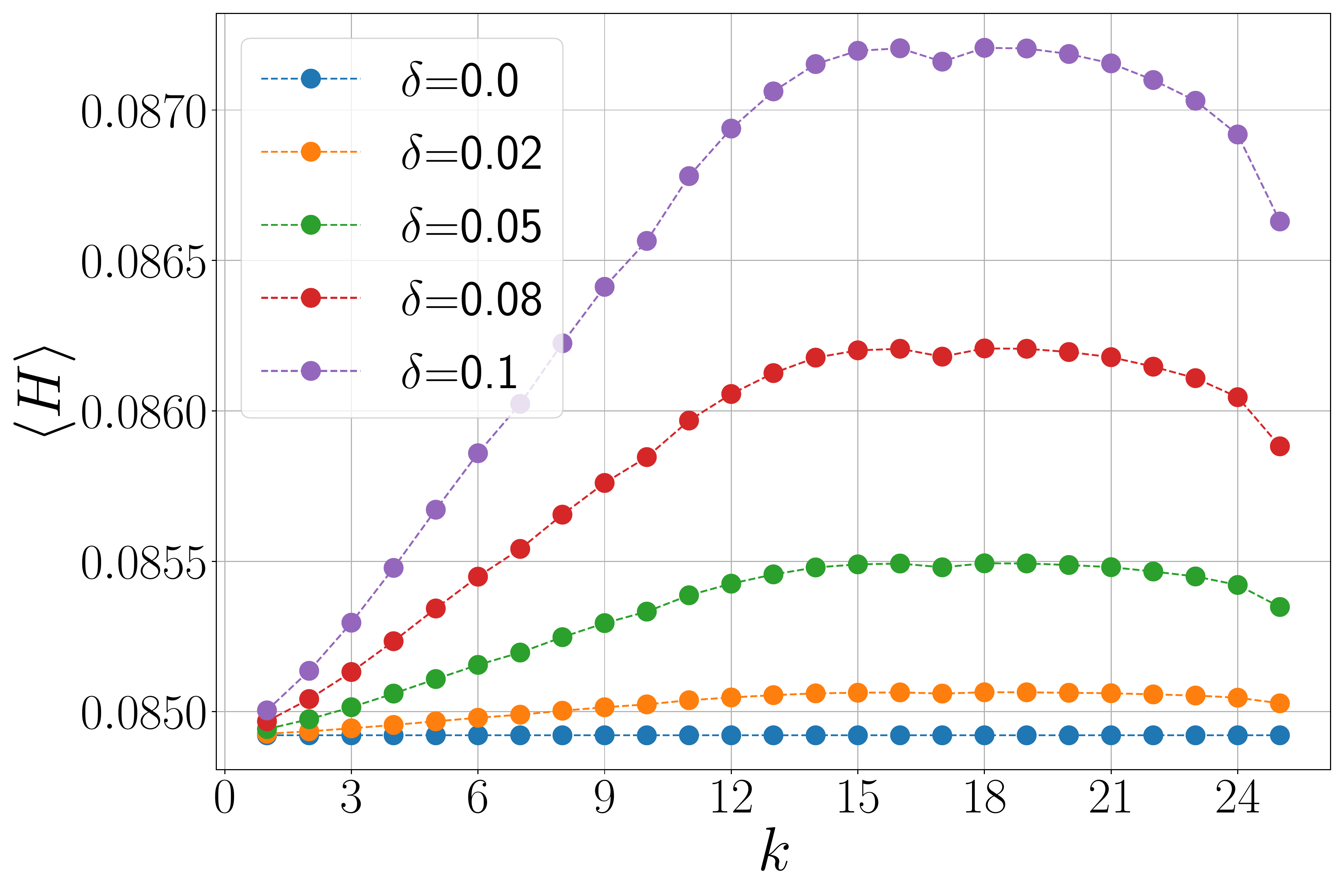}}
   \end{minipage}
   \begin{minipage}[b]{\linewidth}
   \raggedright{~~~\includegraphics[clip=true,width=2.9in]{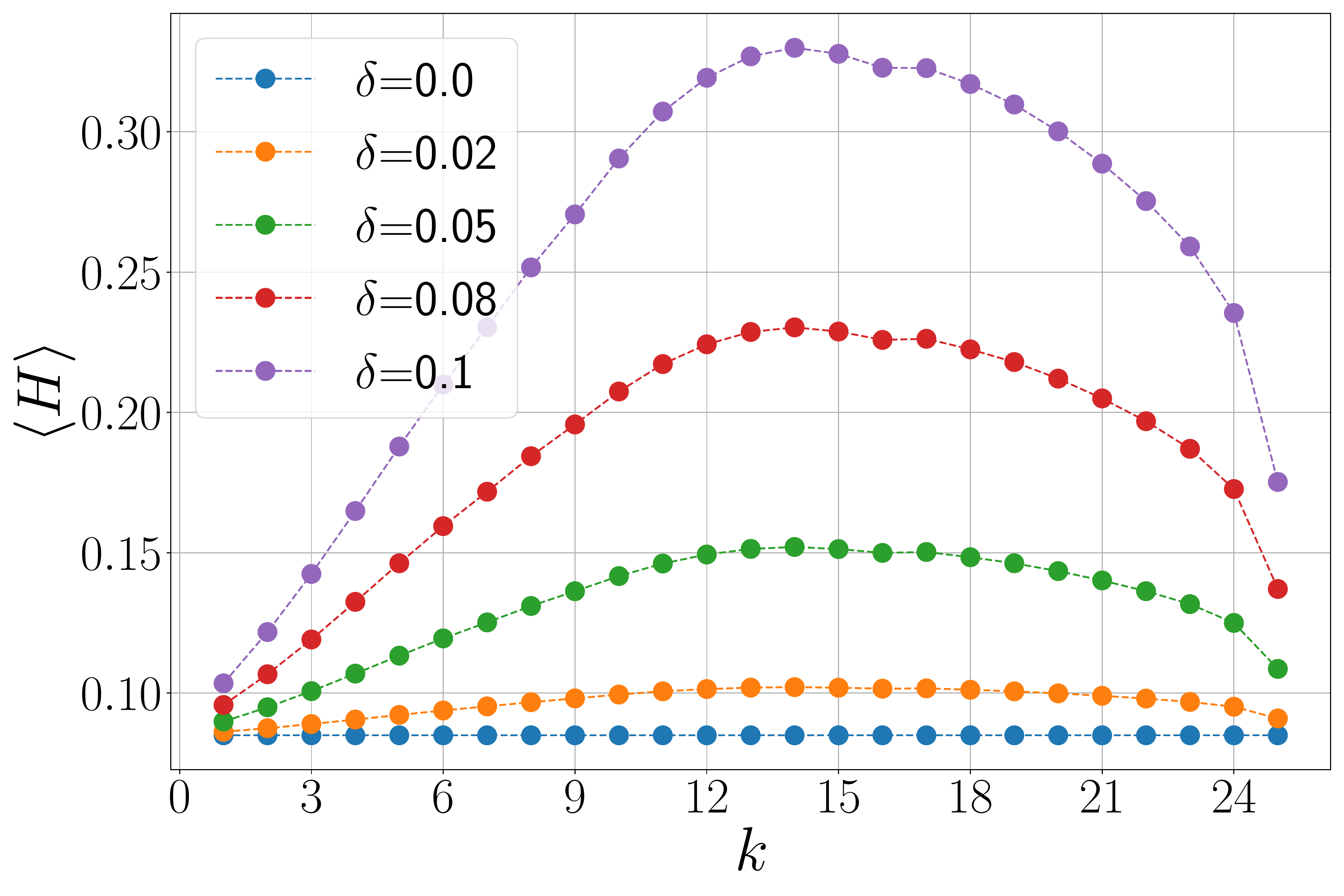}}
   \end{minipage}
   \caption{Energy $\langle H\rangle=\bra{\psi(\bm \theta^*+\bm\delta \bm\theta)}H\ket{\psi(\bm\theta^*+\bm\delta\bm\theta)}$ obtained for unstructured search with $n=10, ~p=25$, when only $\beta_k$ (top) or $\gamma_k$ (bottom) are perturbed by $\delta$.}
 \label{search_energy_individuial_angles}
 \end{figure}

\begin{figure}[!tbh]
 \begin{minipage}[b]{\linewidth}
   \raggedright{~~~\includegraphics[clip=true,width=3.in]{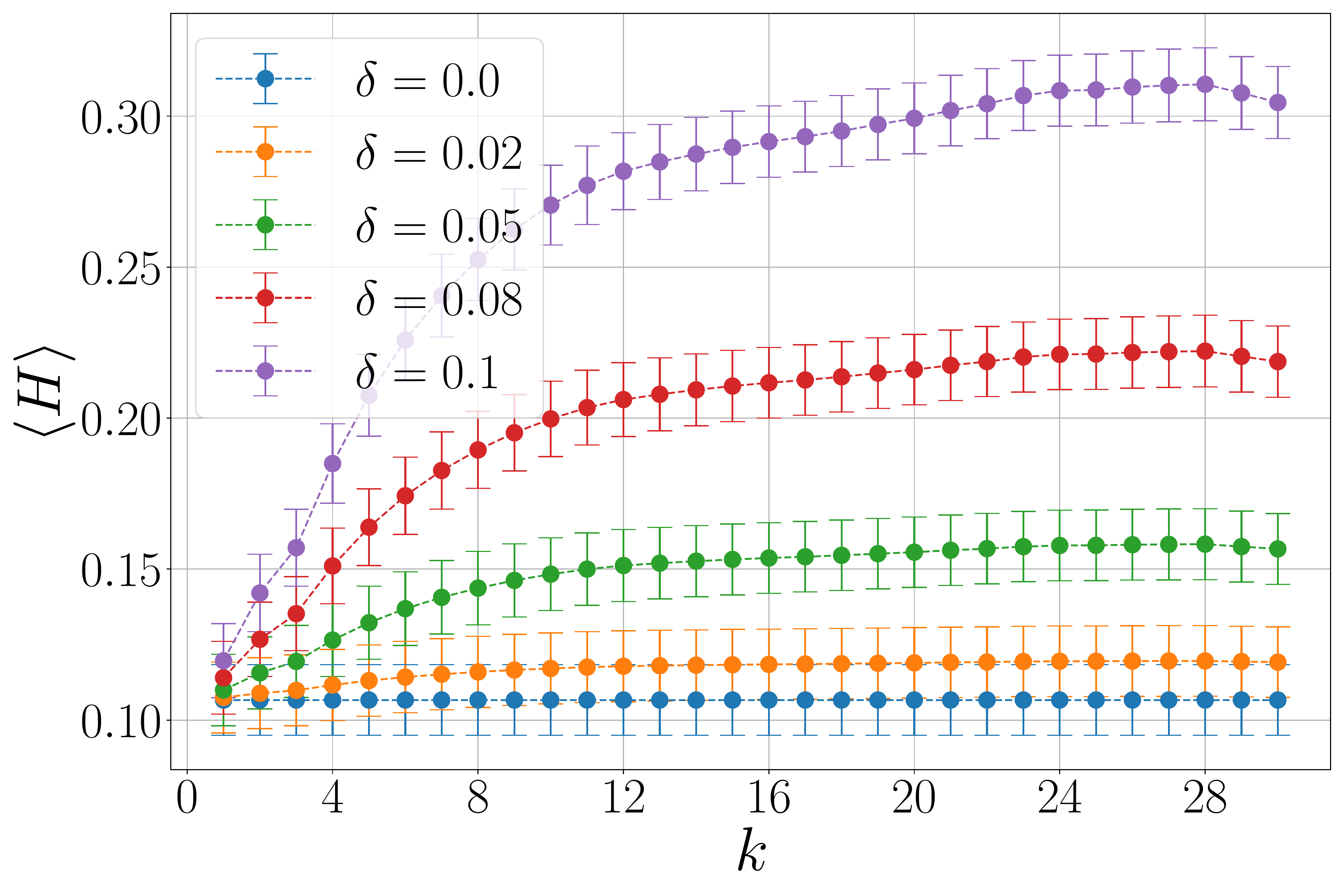}}
   \end{minipage}
   \begin{minipage}[b]{\linewidth}
   \raggedright{~~~\includegraphics[clip=true,width=3.in]{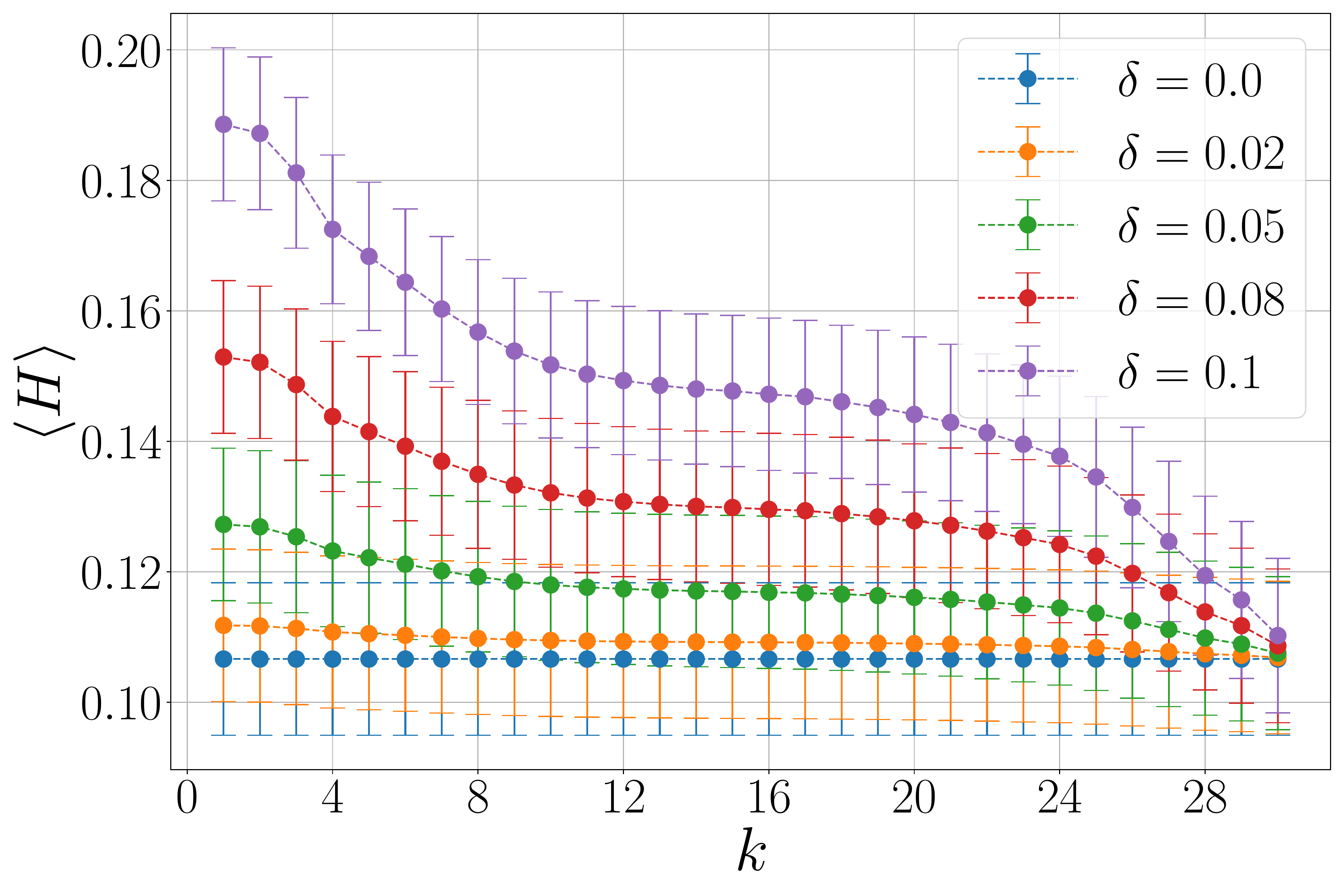}}
   \end{minipage}
   \caption{Average energy $\langle H\rangle=\bra{\psi(\bm \theta^*+\bm\delta \bm\theta)}H\ket{\psi(\bm\theta^*+\bm\delta\bm\theta)}$ of 100 uniformly generated 3-SAT instances solved with $p=30$ layers, where $\beta_k$ (top) or $\gamma_k$ (bottom), from the $k$-th layer, are perturbed by $\delta$.
The instances are of $n=10$ qubits with a clause to variable ratio of 4.2 and a unique satisfying assignment.}
 \label{sat_energy_individuial_angles}
 \end{figure}

In this section we aim at comparing sensitivity of energy to perturbations of specific layers of QAOA circuits. For that we perturb 
the angles $\gamma_k$ and $\beta_k$ in \eqref{ansatz} one at a time by a constant $\delta$, while the rest are kept intact.  Effect of this perturbations on the energy is illustrated in Figures \ref{search_energy_individuial_angles} and \ref{sat_energy_individuial_angles} for $n=10$ qubits, while similar results were also obtained for $n=6, 8$ qubits. These results are numerical and are yet to be explained analytically.  We observe that perturbations to certain angles have a significantly smaller effect on the energy. Thus we can infer that reducing the values of such angles would not have a significant effect on performance. This would allow to reduce the circuit execution time $t_{exec}$ in the implementations where it can be attributed to QAOA angles $t_{exec}\propto\sum_{k=1}^p\beta_k +\gamma_k$. Depending on the physical realisation, decrease of this sum can alternatively be attributed to a change in the amplitude of the applied pulse.
% Alternatively, increasing depth to $p+1$ while limiting the maximum execution time  to that of the original circuit, $ t_{exec}^{p+1} \leq t^{p+1}_{max} = t^p_{exec}$, one can potentially improve performance.
% , since
% \begin{equation}
%      \min\bra{\psi_p} H \ket{\psi_p}\geq \min \bra{\psi_{p+1}} H \ket{\psi_{p+1}}.
% \end{equation}

\begin{figure*}
    \centering
    \includegraphics[width=\textwidth]{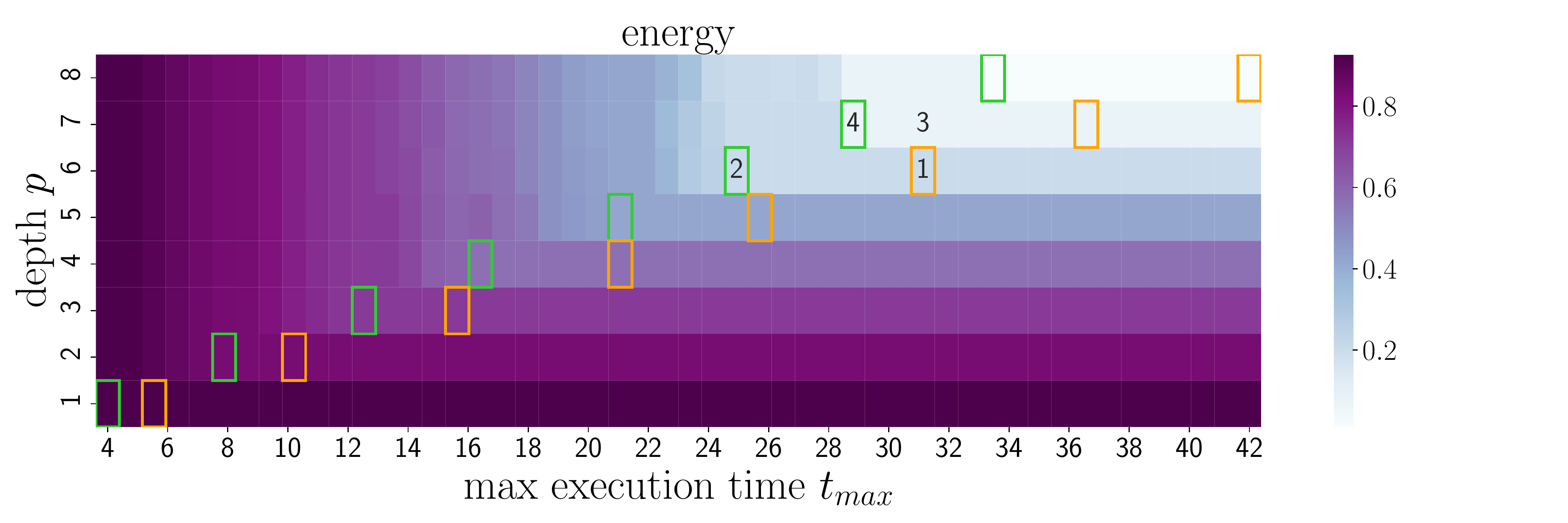}
    \caption{Each cell in the figure represents the optimized energy for a given depth and a fixed maximal execution time. The coloring scheme is indicated on the right. Green and orange rectangles depict the two branches of angles that minimize expectation value for a given depth.}
    \label{exp_values_timelimit}
\end{figure*}

Reducing the execution time is important to quantum algorithms, since variational parameters are proportional to the time required to execute the gates experimentally. NISQ era devices suffer from limited coherence, thus reducing execution times can lead to more efficient hardware utilization \cite{yang2017optimizing,ibrahim2022pulse}.
We test these ideas in the setting of unstructured search, as depicted in Fig.~\ref{exp_values_timelimit}. 
Here we demonstrate QAOA energies for 6 qubits, optimized with an additional condition $t_{exec}\equiv\sum_{k=1}^p\beta_k +\gamma_k\leq t_{max}$ for multiple values of $t_{max}$ and circuit depth $p$. 
This becomes a constrained optimization problem when $t_{max}<3\pi p$ (see ranges for $\gamma_k$ and $\beta_k$).
The highlighted green and orange rectangles correspond to the two groups of optimal angles, obtained in the unconstrained optimization, that minimize the energy at each depth \cite{akshay2021parameter}.

Several interesting conclusions could be drawn from Fig.~\ref{exp_values_timelimit}.
Firstly, it is evident that, for a fixed $p$, starting from a low value of $t_{max}$ and progressively increasing it improves the performance only until the first set of optimal parameters is found (green rectangles).
Past this point performance stagnates; nevertheless, for a sufficiently large value of $t_{max}$ the other solution appears (orange rectangles). 
Thus, for a fixed depth, $t_{exec}$ corresponding to orange boxes is always larger compared to that of the green boxes. Secondly, one may adopt several strategies for reducing execution time without compromising the performance. We illustrate this with a few examples.

Let us consider the labeled rectangles in Fig. \ref{exp_values_timelimit}. It is seen that for rectangle $1$ $t_{exec}\approx 31$ is greater than that of rectangle $2$ which has $t_{exec}\approx25$. Thus by constraining $t_{max}$ to values below $31$, one can force optimization to find solution $2$ instead of solution $1$, keeping $p$ fixed. This significantly reduces the execution time while maintaining a constant energy.  
Furthermore, keeping $t_{max}\approx31$, one can increase the circuit depth from $6$ to $7$, thus going from rectangle $1$ to $3$. In doing so we achieve an improved energy. Note that although $t_{max} \approx 31$ for rectangle $3$, the corresponding $t_{exec} \approx 29$ (same as the execution time of rectangle $4$). Thus by including an extra layer we not only manage to improve the performance, but also reduce the execution time.

% Note that although $t_{max}$ for solution $3$ is the same as that of $1$, its $t_{exec}$ is the same as that of $4$.

% In both the cases the execution time $t_{exec}$ equals the corresponding $t_{max}$.

% Green rectangles also indicate the depth and $t_{exec}$ at which the ansatz will not be able to decrease the energy by either  increasing depth or $t_{max}$. 
% Following the observations of Fig.~ \ref{search_energy_individuial_angles}, by slightly reducing $t_{max}$ the optimizer may reduce the parameters to which the energy is less sensitive. 
% This results in a slight energy increase as illustrated in Fig.~\ref{exp_values_timelimit} where to the left of the green rectangles we can observe darkening gradients.

% By contrast, orange rectangles highlight longer execution times corresponding to different sets of angles that also minimize the energy for a given number of layers. Therefore, if the optimization routine finds the solution corresponding to the orange rectangle, setting $t_{max}$ to be slightly less than the $t_{exec}$ of the orange rectangle will lead the optimizer to find angles corresponding to the green rectangle.  This will amount to a considerable reduction in execution time. 
% Alternatively, increasing the number of layers while keeping $t_{max}$ constant may reduce the energy. 

In general, for an arbitrary problem Hamiltonian one can not be sure if optimization has returned the ideal set of angles (green ones in our example). For this reason, one might employ several strategies based on the behaviour depicted in Fig.~\ref{exp_values_timelimit} to achieve a minimum threshold energy. These include---to reduce $t_{max}$ until energy starts degrading or to increase depth with fixed $t_{max}$ until performance stagnates.

% \begin{widetext}
    
% \onecolumngrid

% \end{widetext}

% \begin{widetext}
    
% \onecolumngrid

% \end{widetext}

% \todo[inline]{Check what happens if not all the angles are independent! $\sigma^2$ behaviour seems to hold anyway...}
% Note that 
% \begin{itemize}
%     \item \eqref{dm_correction} has a very straightforward explanation. It is simply composed of the averaged terms of the circuits where only one angle is perturbed at a time. When the perturbation is small, circuit where two and more angles are perturbed at a time give a quadratically smaller contributions and can be ignored.
%     \item Maybe we can remove several restrictions on the form of perturbations, as \eqref{dm_projector} can be integrated analytically in general. The issue is, for example, that odd powers of $\sin (\delta\theta_k)$ will not vanish unless we request either smallness of perturbation or $p(\delta\theta_k)=p(-\delta\theta_k)$, which might be too artificial.
% \end{itemize}

\section{Discussion}

In this study, we considered a noise model which naturally arises from the control laser pulse intensity fluctuation and spatial inhomogeneity and can become a dominant source of error for certain experimental setups. Within this model the gate parameters receive stochastic perturbation, which leads to an increase of the energy $E^*$ of an optimized variational circuit. Through a perturbative analysis, we found that the change in energy $\delta E$ due to the presence of the gate errors behaves quadratically with respect to the spread of parameter deviations, which is equivalent to linear dependence on the gate error probabilities. Using this result, we derived an upper bound on the level of noise that can be tolerated, satisfying the acceptance condition.

Our analytical findings are substantiated by numerical simulations of VQE for Ising chain with transversed field and QAOA for three common problems - 3-SAT, MAX-CUT and unstructured search - using different modifications of the considered noise model. We confirm that energy perturbation scales quadratically with $\sigma^2$ up to $q\sigma^2\gtrsim1$, expanding applicability of our result compared to analytical expectations. We study a wide range of problem sizes and circuit depths, and show that for realistic values of $\sigma^2\sim10^{-3}-10^{-2}$ the predicted behaviour prevails for circuits up to thousands of gates.
Additionally, our numerical results showed that the algorithmic performance is more resilient to perturbations of certain variational parameters.
Based on this observation, we proposed a strategy to improve performance and reduce the execution time of variational quantum algorithms. Specifically, we showed that the performance of QAOA is not affected when limiting the maximum execution time to $t_{max}=t_{exec}- \epsilon$ for $\epsilon \ll t_{exec}$. We also demonstrated that in some cases reducing $t_{max}$ can lead to significant reductions in $t_{exec}$, while increasing the depth of the algorithm can lead to an energy reduction with fixed $t_{exec}$.

While we presented results for additive perturbation $\theta\to\theta+\delta\theta$, a multiplicative perturbation $\theta\to(1+\epsilon)\theta$ might represent another realistic scenario. Nevertheless, qualitative behaviour for both types of errors was observed to be the same, therefore we focused on the additive scenario in the text. Importantly, whereas our study primarily focused on energy perturbations around the noiseless optimum $\bm\theta^*$, in practice, one has to train the algorithm in the presence of noise, which can change the optimal angles $\bm\theta^*$ to $\bm\theta^*+\bm{\delta\theta}^*$, where the shift $\bm{\delta\theta}^*$ depends on the strength of the noise. However, using perturbation theory around the noiseless optimum,  one can estimate $\bm{\delta\theta}^*=O(\sigma^2)$ in the regime of small noise, and the corresponding change in the energy is ${\rm Tr}(\rho(\bm\theta^*+\bm{\delta\theta}^*)H)- {\rm Tr}(\rho(\bm\theta^*)H)=O(\sigma^4)$.
In other words, even by studying $E(\bm\theta^*)$ we can adequately predict the behavior of true optimal energy $\delta E(\bm\theta^*+\bm{\delta\theta^*})$ . For detailed calculations, please refer to appendix \ref{appen:optimum_shift}.
Indeed, having trained with this type of noise we observe that $\delta E$ was reduced by no more than $10-15\%$ on figures \ref{fig:averaged_pert} and \ref{fig:p=10}. Finally, since $\delta E$ can only be reduced by updating the optimal parameters, the established linear behaviour remains a valid upper bounds on energy perturbation.

\vspace{1em}
\section*{Acknowledgement}
D.R., E.C., S.A., E.P., D.V.~acknowledge support from the research project, Leading Research Center on Quantum Computing (agreement No.~014/20).
\onecolumngrid
\bibliographystyle{unsrt}
\bibliography{references}

\PRLsep

\vspace{3em}
\twocolumngrid

\appendix

\section{k-SAT, MAX-CUT and unstructured search problems}
\label{appen:problems}

\subsection{k-SAT}
Boolean satifyability, or SAT, is the problem of determining weather a boolean formula written in conjunctive normal form (CNF) is satisfiable. 
It is possible to map any SAT instance via Karp reduction into $k$-SAT, which are restricted to $k$ literals per clause. In order to approximate solutions to SAT we embed the instance into a Hamiltonian as
\begin{equation}
    H_{\text{SAT}}=\sum_j P(j),
\end{equation}
where $j$ indexes clauses of an instance, and $P(j)$ is the tensor product of projectors that penalizes bit string assignments that do not satisfy the $j$-th clause. In this work we focused on the $k=3$ case.

\subsection{MAX-CUT}
Given a graph $(V,E)$, MAX-CUT is a problem of identifying the maximal cut of the graph, i.e. finding a graph bipartition with maximal number of edges between two components. Any instance of MAX-CUT problem can be encoded as an $n=|V|$ qubit problem Hamiltonian
\begin{equation}
    H = \sum_{(i,j) \in E} Z_i Z_j,
\end{equation}
whose ground state encodes the solution to the problem: bit sting $\ket{z_1z_2,\dots,z_n}$ corresponds to the partition where $i$-th vertex belongs to the $z_i=0,1$ component. 

\subsection{Unstructured search}
Consider an unstructured database $S$ indexed by  $j\in\{0,1\}^{\times n}$. Let $f: \{0,1\}^{\times n} \rightarrow \{0,1\}$ be a Boolean function (a.k.a. black box) such that:
\begin{equation}
    f(j) = \begin{cases}
    1 & \text{iff} ~ j=t \\
    0 & otherwise.
    \end{cases}
\end{equation}
The task is to find $t \in \{0,1\}^{\times n}$.
The corresponding problem Hamiltonian for QAOA is 
\begin{equation}
    H_t=\mathbb{1}-\ketbra{t}{t},
\end{equation}
thus the expected value is given by
\begin{equation}
    \langle H\rangle = 1-|\braket{t}{\psi_p(\bm\gamma,\bm\beta)}|^2.
\end{equation}
QAOA performance for unstructured search is not sensitive to the particular target state $\ket{t}$ in the computational basis. 
For any target state $\ket{t}$ representing a binary string, there is a $U=U^\dagger$ composed of $X$ and $\mathbb{1}$ operators such that $U\ket{0}^{\otimes n}=\ket{t}$.
The overlap of an arbitrary state prepared by a QAOA sequence with $\ket{t}$ is then:
\begin{align*}
    \braket{t}{\psi_p(\bm\gamma,\bm\beta)} &= \bra{t} \prod\limits_{k=1}^p e^{-i \beta_k \mathcal{H}_{x}} e^{-i \gamma_k \ketbra{\bm{t}}{\bm{t}}}\ket{+}^{\otimes{n}}\\
    &= \bra{0}^{\otimes n}U \prod\limits_{k=1}^p e^{-i \beta_k \mathcal{H}_{x}} e^{-i \gamma_k U(\ketbra{0}{0})^{\otimes n}U}\ket{+}^{\otimes{n}}\\
    &=\bra{0}^{\otimes n}U \prod\limits_{k=1}^p e^{-i \beta_k \mathcal{H}_{x}} Ue^{-i \gamma_k (\ketbra{0}{0})^{\otimes n}}U\ket{+}^{\otimes{n}}\\
    &=\bra{0}^{\otimes n} \prod\limits_{k=1}^p e^{-i \beta_k \mathcal{H}_{x}} e^{-i \gamma_k (\ketbra{0}{0})^{\otimes n}}\ket{+}^{\otimes{n}},
\end{align*}
which is independent on $t$.

\section{QAOA for MAX-CUT and unstructured search}
\label{appen:other_results}
\subsection{MAX-CUT}
\begin{figure}[ht!]
    \centering
    \includegraphics[width=0.49\textwidth]{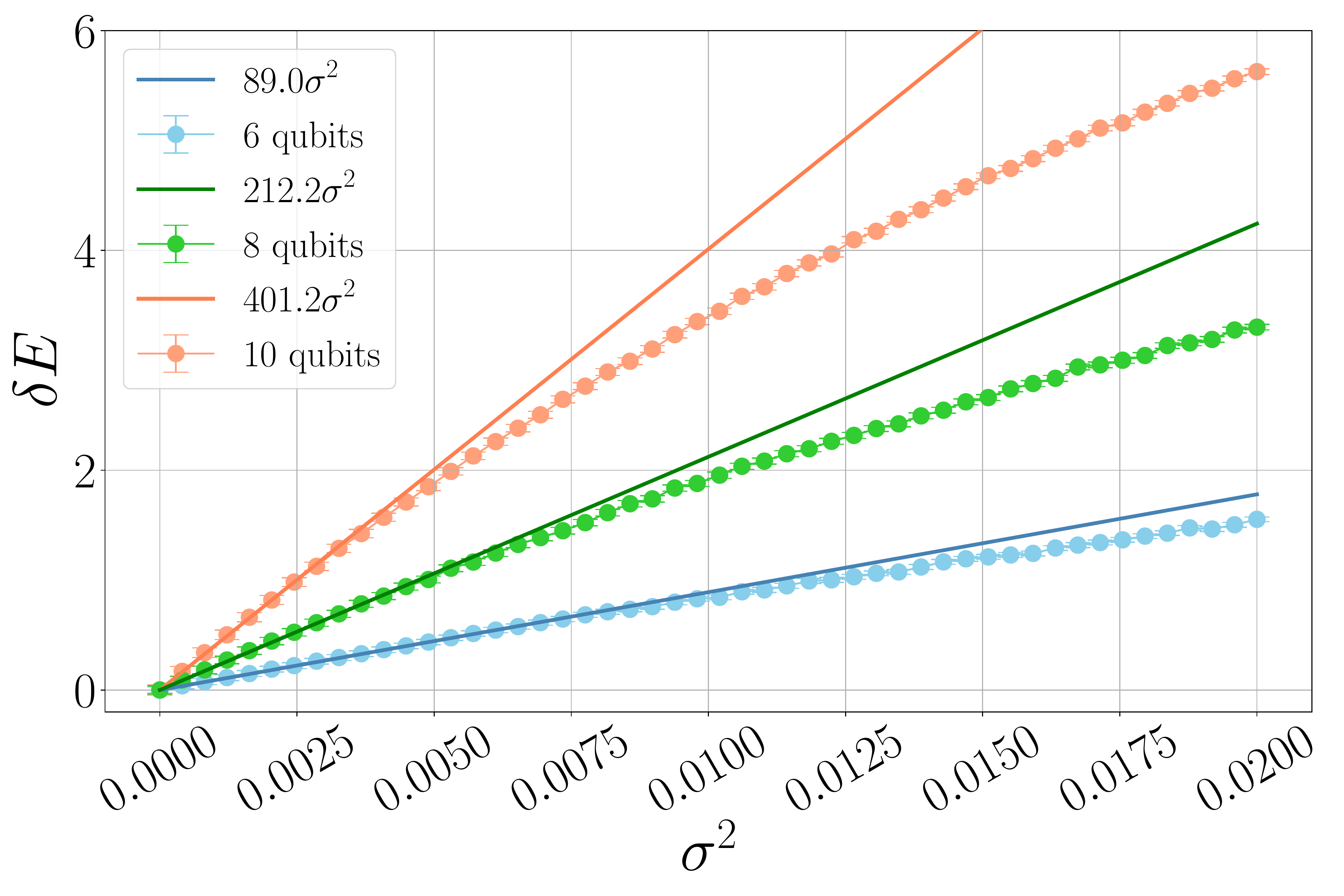}
    \caption{Average energy shift of 100 instances of of MAX-CUT on 6, 8 and 10 qubits obtained from perturbed optimal circuits of depth $10$, $13$ and $15$, respectively. The instances were designed to approximately have graph density $0.5$. 
    Error bars depict standard errors. Polynomial fits of data indicate $\delta E\propto \sigma^2$ up to $q\sigma^2\gtrsim1$.}
    \label{fig:max_cut}
\end{figure}

Here we show results of perturbing QAOA circuits, optimized to solve instances of MAX-CUT. We follow the same procedure explained in the main text. We consider $100$ randomly generated graphs of $n=6, 8, 10$ vertices; these graphs were generated by adding edges between every pair of vertices with probability $1/2$. The instances were minimized with circuits of depth $10$, $13$ and $15$, respectively, which guaranteed at least $85\%$ overlap with the ground states. The energies of perturbed circuits are illustrated in Fig. \ref{fig:max_cut}.
Similar to the results of Section \ref{sec:results} our linear prediction holds up to $q\sigma^2\gtrsim1$.

\subsection{Unstructured search}
\begin{figure}[ht!]
    \centering
    \includegraphics[width=0.49\textwidth]{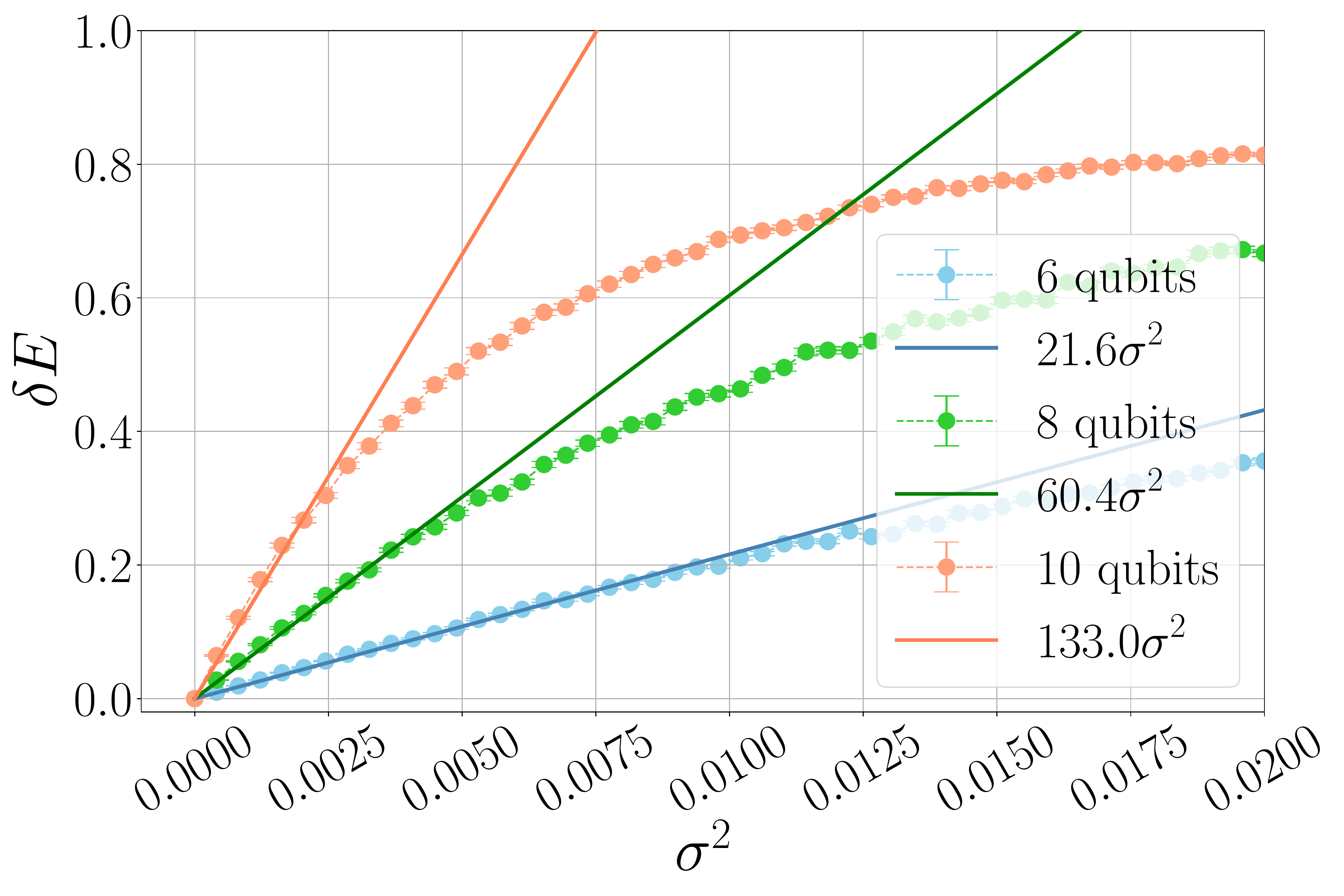}
    \caption{Average energy for the problem of unstructured search obtained by the perturbation of $\bm \gamma^*,~ \bm\beta^*$ by $\delta$ uniformly sampled from the range $(-\sigma,\sigma)$. Error bars depict standard errors. Polynomial fits of data points of 6, 8 and 10 qubits in the ranges $\sigma \in [0,0.1], ~[0,0.07], ~[0,0.05]$, respectively confirm that $\delta E \propto \sigma^2$.}
    \label{fig:state_prep_sigma}
\end{figure}

Here we show results of perturbing QAOA circuits, optimized to solve the unstructured search problem. Opposite to the main text, here we consider perturbations introduced to every QAOA layer, without decomposition into gates of the form \eqref{eq:op}. This corresponds to multiple gates receiving the same perturbation in the settings presented in Section \ref{sec:results}. Energy increases of the perturbed optimized circuits are illustrated in Fig. \ref{fig:state_prep_sigma}. Despite a different setting where problem Hamiltonian propagator is not decomposed into gates of form \eqref{eq:op} the energy perturbation still scales linearly with $\sigma^2$ at least in the range of small $\sigma$, which demonstrates that our result applies even to circuits of a more general type. 

% \section{Energy variation in presence of constant perturbations to gate parameters}
% \label{appen:constant_shift}
% Using \eqref{eq:dm_projector} one can calculate perturbation to the energy caused by a shift of the optimal angles by a constant $\bm\delta \bm\theta$ as
% \begin{align}
%     \delta E&=\bra{\psi(\bm \theta^*+\bm\delta \bm\theta)}H\ket{\psi(\bm\theta^*+\bm\delta\bm\theta)}-\bra{\psi(\bm\theta^*)}H\ket{\psi(\bm\theta^*)}\nonumber \\&=
%     -\sum_{k=1}^q\delta\theta_k^2E^*+\sum_{m\ne k}^q\delta\theta_k\delta\theta_m(\bra{\psi(\bm\theta^*)}H\ket{\psi_{km}}+h.c.)\nonumber\\&~~~~+ \sum_{m,k}^q\delta\theta_k\delta\theta_k\bra{\psi_m}H\ket{\psi_k} + o(\delta\theta_k\delta\theta_m)\nonumber\\&=
%     \dfrac{1}{2}(\bm{\delta\theta})^T\bm H \bm{\delta\theta}+o(\delta\theta_k\delta\theta_m),
% \end{align}

% where $\ket{\psi_{mk}}=\ket{\psi_{0...1...1...0}}$ with $1$ placed only at $m$-th and $k$-th positions.  $\bm H$ is the Hessian of the energy at noiseless optimum, $\bm H_{ij}=\dfrac{\partial^2}{\partial\theta_i\partial\theta_j}\bra{\psi(\bm\theta)}H\ket{\psi(\bm\theta)}\vert_{\bm\theta=\bm\theta^*}$. Here we use the fact that at the optimal position linear contribution to the cost function necessarily vanishes. 
% It is seen now that for the constant perturbation $\delta\theta_k=\delta$ the energy changes as $\delta E\propto\delta^2$.

\section{Shift of optimal parameters in the presence of noise}
\label{appen:optimum_shift}
Let us use expressions \eqref{eq:rho} and \eqref{eq:delta_rho} to estimate change in the energy if one accounts for shift of optimal parameters $\bm\theta^*\to\bm\theta^*+\bm{\delta\theta}^*$ in the presence of noise:
\begin{align}
    &{\rm Tr}(\rho(\bm\theta^*+\bm{\delta\theta}^*)H)=\nonumber\\
    &~~~(1-\sum_{k=1}^qa_k)\bra{\psi(\bm\theta^*+\bm{\delta\theta}^*)}H\ket{\psi(\bm\theta^*+\bm{\delta\theta}^*)}\nonumber\\
    &+\sum_{k=1}^qa_k\bra{\psi_k(\bm\theta^*+\bm{\delta\theta}^*)}H\ket{\psi_k(\bm\theta^*+\bm{\delta\theta}^*)}+o(\sigma_k^2)
\end{align}

We introduce gradients of the noisy terms 

\begin{eqnarray}
\bm B^k=\dfrac{\partial}{\partial\bm \theta}\bra{\psi_k(\bm\theta)}H\ket{\psi_k(\bm\theta)}\vert_{\bm\theta=\bm\theta^*},
\end{eqnarray}
and Hessian $\bm H$
of the energy at noiseless optimum, 
\begin{equation}
\bm H_{ij}=\dfrac{\partial^2}{\partial\theta_i\partial\theta_j}\bra{\psi(\bm\theta)}H\ket{\psi(\bm\theta)}\vert_{\bm\theta=\bm\theta^*}.
\end{equation}
Notice that gradients of the noiseless function $\bra{\psi(\bm\theta)}H\ket{\psi(\bm\theta)}$ vanish at optimum. Then,
\begin{align}
    &{\rm Tr}(\rho(\bm\theta^*+\bm{\delta\theta}^*)H)\approx(1-\sum_{k=1}^qa_k)E^*+\dfrac{1}{2}(\bm{\delta\theta}^*)^T\bm H \bm{\delta\theta}^*\nonumber\\
    &+\sum_{k=1}^q a_k[\bra{\psi_k(\bm\theta^*)}H\ket{\psi_k(\bm\theta^*)}+(\bm{\delta\theta}^*)^T \bm B^k].
\end{align}

Minimizing it with respect to $\bm{\delta\theta}^*$ one gets
\begin{equation}
   \bm{\delta\theta}^* = -\sum_{k=1}^q a_k \bm H^{-1}\bm B^k.  
\end{equation}
Thus, if we account for the change of optimal parameters in the presence of noise, the energy shifts by 
\begin{align}
    &{\rm Tr}(\rho(\bm\theta^*+\bm{\delta\theta}^*)H)- {\rm Tr}(\rho(\bm\theta^*)H)\approx\nonumber\\
    &(\bm{\delta\theta}^*)^T\bm H \bm{\delta\theta}^*
    +\sum_{k=1}^q a_k(\bm{\delta\theta}^*)^T \bm B^k
    =O(\sigma^4).
\end{align}
\end{document}